\documentclass[12pt]{article}

\usepackage{amsmath}
\usepackage{amssymb}
\usepackage{graphicx}
\usepackage{float}



\begin{document}
\title{Oscillations in $p$-adic diffusion processes and simulation of the conformational dynamics of
protein }
\author{A.\,Kh.~Bikulov \medskip{}
 \\
 \textit{Institute of Chemical Physics, } \\
 \textit{Kosygina Street 4, 117734 Moscow, Russia} \medskip{}
 \\
 e-mail:\:\texttt{beecul@mail.ru} \bigskip{}
 \\
 and \bigskip{}
 \\
 A.\,P.~Zubarev \medskip{}
 \\
 \textit{ Physics Department, Samara University, } \\
 \textit{ Moskovskoe shosse 34, 443123, Samara, Russia } \medskip{}
 \\
 \textit{Natural Science Department, } \\
 \textit{Samara State University of Railway Transport,} \\
 \textit{Perviy Bezimyaniy pereulok 18, 443066, Samara, Russia }\medskip{}
 \\
 e-mail:\:\texttt{apzubarev@mail.ru} }
\maketitle
\begin{abstract}
Logarithmic oscillations superimposed on a power-law trend appear
in the behavior of various complex hierarchical systems. In this
paper, we study the logarithmic oscillations of relaxation curves
in $p$-adic diffusion models that are used to describe the
conformational dynamics of protein. We consider the case of a
purely $p$-adic diffusion, as well as the case of $p$-adic
diffusion with a reaction sink. We show that, relaxation curves
for large times in these two cases are described by a power law on
which logarithmic oscillations are superimposed whose period and
amplitude are determined by the parameters of the model. We also
provide a physical explanation of the emergence of oscillations in
relaxation curves and discuss the relation of the results to the
experiments on relaxation dynamics of protein.

\vspace{5mm}
\end{abstract}

\section{Introduction}

If a dynamical variable $x\left(t\right)$ of a physical system
behaves as

\begin{equation}
x\left(t\right)=At^{-\beta}f\left(\log t\right)\label{log_per}
\end{equation}
in some time domain, where $A,\:\beta\:\in\:\mathbb{R}$ and
$f\left(y\right)$ is a periodic function, then it is said that
logarithmic oscillations superimposed on a power-law trend occur
in the system. In the early 1990s a number of authors (see, for
example, \cite{Blatz,Anifrani,Sahimi,Johansen_1}) found that the
time variation of the dynamical variables of some physical,
chemical, and biological systems is described by a law similar to
(\ref{log_per}). In a series of papers devoted to the study of the
dynamics of financial markets (see, for example,
\cite{Feigenbaum,Gluzman,Drozdz,Sornette} and references therein),
the authors also found through the analysis of empirical data that
the behavior of the logarithms of prices in financial markets
before crashes also resembles a power law multiplied by a sum of
log-periodic harmonics, where the presence of logarithmic
oscillations is associated with the discrete scale invariance of
the system \cite{Saleur_1,Saleur_2}. In the simplest case, this
connection can be traced as follows. Consider a real function
$x\left(t\right)$ of time $t$ that describes a dynamical variable
of the system. The discrete scale invariance of the function
$x\left(t\right)$ implies that there exists a
$\theta\in\mathbb{R}_{+}$ such that the following relation holds
for some $\beta\in\mathbb{R}$:
\begin{equation}
x\left(t\right)=\theta^{\beta}x\left(\theta t\right).\label{si}
\end{equation}
The general solution of equation (\ref{si}) is given by
(\ref{log_per}), where $f\left(y\right)$ is a periodic function
with period $\log\theta$. If relation (\ref{si}) holds
asymptotically as $t\rightarrow\infty$ rather than exactly, then
the law (\ref{log_per}) also holds asymptotically.

The discrete scale invariance of a dynamical variable of a system
implies a self-similar dynamics of the system at different time
scales. Such a scenario can be realized for a hierarchical
self-similar organization of the state space of the system. These
systems include, in particular, systems whose state space is
ultrametric. In \cite{MKJ} the authors argued that the relaxation
of a system in the ultrametric space described by the boundary of
a Bethe tree should lead to the dynamics of observables that can
be represented as a sum of exponential functions of time $t$ and
this sum can be represented as the product of a power function of
$t$ multiplied by a periodic function of $\log t$.

In physics, the models of real systems in ultrametric spaces have
first been considered in the 1970s in relation to the study of
disordered spin systems (spin glasses)
\cite{RTV,Dayson,Parisi1,Dot}. Almost simultaneously with the
appearance of ultrametric models of spin glasses, it was suggested
that the space of conformational states of a protein molecule is
ultrametric \cite{Frauen1,Frauen2}. A significant contribution to
the development of ultrametric (and the related $p$-adic)
description of complex systems of different nature, including
protein, has been made in the last 30 years (for a review, see
\cite{ALL,ALL_1,ALL_2}). A detailed validation of the ultrametric
approach to the simulation of the conformational dynamics of
protein is given in a series of papers of the present authors
\cite{ABK_1999,ABKO_2002,ABO_2003,ABO_2004,AB_2008,ABZ_2009,ABZ_2011,ABZ_2013,ABZ_2014,BZ_2021}
and can be briefly summarized as follows (see \cite{BZ_2021} for
more details). At a given temperature of the medium, protein
executes thermal motion, which represents a random walk on the
configuration space of the degrees of freedom of a macromolecule.
To describe this random walk of protein within the Langevin or the
Fokker--Planck approach, one should precisely define all degrees
of freedom and the potential energy of protein, which is hardly
possible. Therefore, one applies the following approximation to
describe the conformational dynamics of protein. The configuration
space of protein is divided into subsets -- elementary basins
(attraction basins). Each such elementary basin is associated with
a local minimum of the potential energy of protein and is defined
as an open subset of points from which one can reach the local
minimum by gradient descent (see \cite{Stillinger1,Stillinger2,BK}
for details). On the set of elementary basins, one can introduce a
distance function (a metric), which is defined for any two
elementary basins in terms of the minimum potential barrier over
all paths connecting the configuration minima of these basins. One
can show that this metric is an ultrametric, and thus the set of
elementary basins is an ultrametric space \cite{Koz}. Elementary
basins can be combined into sets, which are ultrametric balls, and
these sets are also called basins. The conformational state of
protein corresponding to a certain basin is a quasi-equilibrium
macrostate in which a random walk of protein on a given basin with
a distribution function close to the equilibrium distribution
function in this basin is implemented. Thus, the set of
conformational states associated with elementary basins is also an
ultrametric space. One of simplifying assumptions is that this
ultrametric space is homogeneous. This allows one to perform a
$p$-adic parameterization of the space of conformational states of
protein, i.e., to map this space to the field of $p$-adic numbers
$\mathbb{Q}_{p}$ (see \cite{VVZ} for the introduction to $p$-adic
analysis). Namely, the set of all possible conformational states
of protein corresponding to all possible basins is parameterized
by a set of $p$-adic balls of radius greater than or equal to $1$.
In this case, the dynamics of protein on the set of conformational
states is represented by a random walk on the field
$\mathbb{Q}_{p}$, which is described by a Markov random process
$\xi\left(t,\omega\right)\colon\:\mathbb{R}\times\Omega\rightarrow\mathbb{Q}_{p}$.
The density of the distribution function $f\left(x,t\right)$ of
such a process is assumed to be a locally constant function with
radius of local constancy equal to one (i.e., for any $x$ and
$x'$, $|x'|_{p}\leq1$, the equality
$f\left(x\right)=f\left(x+x'\right)$) holds, and this function is
a solution of the equation of $p$-adic random walk (the
Kolmogorov--Feller equation in the field of $p$-adic numbers),
\begin{equation}
\frac{\partial f(x,t)}{dt}=\intop_{\mathbb{Q}_{p}}W\left(|x-y|_{p}\right)\left(f\left(y,t\right)-f\left(x,t\right)\right)dy,\label{EQ_URW}
\end{equation}
in which the kernel $W\left(|x-y|_{p}\right)$ of the integral
operator depends only on the ultrametric distance $|x-y|_{p}$
between points $x$ and $y$. Equation (\ref{EQ_URW}) is assumed to
be covariant under the scale transformations $x\rightarrow
x^{\prime}=\lambda x$ and $t\rightarrow
t^{\prime}=\left|\lambda\right|_{p}^{\alpha}t$, where
$\lambda\in\mathbb{Q}_{p}$ is the transformation parameter and
$\alpha\in\mathbb{R}_{+}$. The last assumption imposes a strong
constraint on the choice of the kernel $W\left(|x-y|_{p}\right)$
of the integral operator in equation (\ref{EQ_URW}). Namely, under
this condition, the kernel of the operator coincides up to a
factor with the kernel of the Vladimirov operator \cite{VVZ}:
\begin{equation}
W\left(|x-y|_{p}\right)\sim\dfrac{1}{|x-y|_{p}^{\alpha+1}}.\label{W}
\end{equation}
In this case, equation (\ref{EQ_URW}) is conventionally called a
$p$-adic diffusion equation, since the integral operator on its
right-hand side is the operator of $p$-adic fractional
differentiation of order $\alpha$. The parameter $\alpha$ can be
assigned a physical meaning if one sets $\alpha=\dfrac{E_{0}}{kT}$
and writes
$\dfrac{1}{|x-y|_{p}^{\alpha}}=\exp\left(-\dfrac{E_{0}\log\left(|x-y|_{p}\right)}{kT}\right),$
where $T$ is temperature, $k$ is the Boltzmann constant, and
$E_{0}$ is a parameter with the dimension of energy. The last
expression is interpreted as a Boltzmann factor defining the
probability that the system overcomes the potential barrier
\begin{equation}
E\left(x,y\right)=E_{0}\log|x-y|_{p}\label{E(x,y)}
\end{equation}
between two conformations containing the points $x$ and $y$,
respectively. In this case, the additional factor
$\dfrac{1}{|x-y|_{p}}$ in (\ref{W}) is inversely proportional to
the combinatorial factor equal to the number of conformations
separated by the potential barrier (\ref{E(x,y)}) from the
conformation containing the point $x$.

Despite the apparent simplicity, the $p$-adic model of
conformational dynamics of protein provided a relevant description
of two main experiments on the relaxation dynamics of protein.
Namely, in \cite{AB_2008,ABZ_2014,BZ_2022} we showed that the
relaxation curves of spectral diffusion experiments
\cite{PFVBB,Friedrich1,Friedrich2} can be described within the
$p$-adic diffusion model. In \cite{ABKO_2002}, we showed that the
relaxation curves of experiments on the kinetics of CO binding to
myoglobin \cite{ABB,SAB} at high temperatures can be described by
the model of $p$-adic diffusion with a reaction sink. Later, in
\cite{ABZ_2013,ABZ_2014,BZ_2021}, we obtained these results for
the entire temperature range.

A characteristic feature of relaxation curves in $p$-adic models
of conformational dynamics of protein is the fact that, at large
times, they can take the form of a power law on which oscillations
are superimposed. Numerical investigations of the solutions of the
$p$-adic diffusion equation on compact subsets of the set
$\mathbb{Q}_{p}$ have shown that the period of these oscillations
increases with time and their amplitude increases with a decrease
in temperature (i.e., with an increase in the parameter $\alpha$).
Nevertheless, a detailed analytical investigation of these
oscillations in the models of $p$-adic diffusion has not been
carried out yet. In the present paper, we eliminate this gap.
Namely, we analyze in detail these oscillations in two cases: in
the case of purely $p$-adic diffusion and in the case of $p$-adic
diffusion with a reaction sink. We show that, in these cases, the
relaxation curves at large times are described by a power law on
which logarithmic oscillations are superimposed whose period and
amplitude are determined by the parameters $p$ and $\alpha$.

The paper is organized as follows. In Section 2 we study the
oscillations of the probability measure of the support of the
initial distribution of the solution to the Cauchy problem for a
$p$-adic diffusion equation with the initial condition on a
compact set. In Section 3 we analyze the oscillations of the
probability measure of the whole $\mathbb{Q}_{p}$ for the solution
of the Cauchy problem for the equation of $p$-adic diffusion with
a reaction sink and with the initial condition on a compact set.
In the concluding section we provide a physical explanation for
the emergence of oscillations on relaxation curves and discuss the
relation of the results obtained to the experiments on the
relaxation dynamics of protein. The proofs of a number of
propositions necessary for Sections 2 and 3 are given in
Appendices 1 -- 3.

\section{Oscillations in the solution of the $p$-adic diffusion equation }

The $p$-adic diffusion equation (see \cite{VVZ}) has the form

\begin{equation}
\dfrac{df\left(x,t\right)}{dt}=-\dfrac{1}{\Gamma_{p}\left(-\alpha\right)}\intop_{\mathbb{Q}_{p}}d_{p}y\dfrac{f\left(y,t\right)-f\left(x,t\right)}{\left|x-y\right|_{p}^{\alpha+1}}.\label{Vlad}
\end{equation}
where
$\Gamma_{p}\left(-\alpha\right)=\dfrac{1-p^{-\alpha-1}}{1-p^{\alpha}}$
is the $p$-adic gamma function. We consider the Cauchy problem for
equation (\ref{Vlad}) with the initial condition on some compact
set chosen in the form of a ring of $p$-adic integers
$\mathbb{Z}_{p}$:
\begin{equation}
f\left(x,0\right)=\Omega\left(\left|x\right|_{p}\right).\label{ic}
\end{equation}
In physical applications, equation (\ref{Vlad}) describes a random
walk on $\mathbb{Q}_{p}$, and $f\left(x,t\right)$ is the density
of the distribution function of the trajectory at time $t$. Of
special interest is to find the so-called function of survival
probability
$S_{\mathbb{Z}_{p}}\left(t\right)\equiv\intop_{\mathbb{Q}_{p}}d_{p}x\Omega\left(\left|x\right|_{p}\right)f\left(x,t\right)$,
which is the probability measure of the support of the initial
distribution at time $t$.

\begin{figure}[H]
\centering{}\includegraphics[scale=0.6]{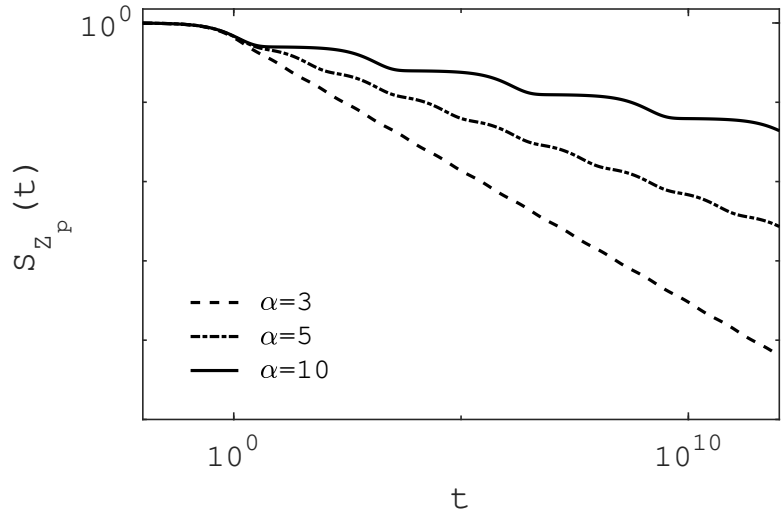}\caption{Graph of the
survival probability
$S_{\mathbb{Z}_{p}}\left(t\right)\equiv\intop_{\mathbb{Z}_{p}}d_{p}xf\left(x,t\right)$,
where $f\left(x,t\right)$ is the numerical solution of equation
(\ref{Vlad}) on the compact set $B_{r}$ with the initial condition
(\ref{ic}) for $p=2$ and $r=20$. \label{Fig_1}}
\end{figure}

One can analyze the Cauchy problem (\ref{Vlad})--(\ref{ic})
numerically by replacing the integration domain $\mathbb{Q}_{p}$
by a $p$-adic ball $B_{r}=\left\{ x:\;\left|x\right|_{p}\leq
p^{r}\right\} $. In this case, equation (\ref{Vlad}) is equivalent
to a system of $r+1$ linear differential equations. Figure
\ref{Fig_1} demonstrates the graphs of the function
$S_{\mathbb{Z}_{p}}\left(t\right)$ in a double logarithmic scale
for the following values of the parameters: $p=2$, $r=20$, and
$\alpha=3,\:5,\:10$. One can see that at large times the curves
represent a power-law function on which oscillations are
superimposed; as $\alpha$ increases, the period and amplitude of
the oscillations increase.

Analytically, the solution of the Cauchy problem (\ref{Vlad}) --
(\ref{ic}) is represented as (see, for example, \cite{ABKO_2002}):
\[
f\left(x,t\right)=\Omega\left(\left|x\right|_{p}\right)\left(1-p^{-1}\right)\sum_{i=0}^{\infty}p^{-i}\exp\left(-p^{-\alpha i}t\right)
\]
\begin{equation}
+\left(1-\Omega\left(\left|x\right|_{p}\right)\right)\left|x\right|_{p}^{-1}\left(\left(1-p^{-1}\right)\sum_{i=0}^{\infty}p^{-i}\exp\left(-p^{-\alpha i}\left|x\right|_{p}^{-\alpha}t\right)-\exp\left(-p^{\alpha}\left|x\right|_{p}^{-\alpha}t\right)\right).\label{f(x,t)}
\end{equation}
In this case, the survival probability is

\begin{equation}
S_{\mathbb{Z}_{p}}\left(t\right)=\left.f\left(x,t\right)\right|_{\left|x\right|_{p}\leq1}=\left(1-p^{-1}\right)\sum_{i=0}^{\infty}p^{-i}\exp\left(-p^{-\alpha i}t\right).\label{S(t)}
\end{equation}
It follows from the form of the solution (\ref{S(t)}) that
\begin{equation}
S_{\mathbb{Z}_{p}}\left(p^{\alpha}t\right)=p^{-1}S_{\mathbb{Z}_{p}}\left(t\right)+\left(1-p^{-1}\right)\exp\left(-p^{\alpha}t\right).\label{Pre_si}
\end{equation}
Since $\exp\left(-p^{\alpha}t\right)$ decreases as
$t\rightarrow\infty$ faster than (\ref{S(t)}), relation
(\ref{Pre_si}) implies the asymptotic discrete scale invariance of
the function $S_{\mathbb{Z}_{p}}\left(t\right)$:

\[
pS_{\mathbb{Z}_{p}}\left(p^{\alpha}t\right)=S_{\mathbb{Z}_{p}}\left(t\right)\left(1+o\left(1\right)\right),
\]
where $o\left(1\right)$ is an infinitesimal function as
$t\rightarrow\infty$. Hence we can make a conclusion about the
general asymptotic behavior of $S\left(t\right)$:

\begin{equation}
S_{\mathbb{Z}_{p}}\left(t\right)\sim t^{-\tfrac{1}{\alpha}}f\left(\log t\right)\left(1+o\left(1\right)\right),\label{as_S}
\end{equation}
where $f\left(x\right)$ is a periodic function with period
$\alpha\log p$. The result (\ref{as_S}) can be obtained rigorously
after determining the explicit form of the function
$f\left(x\right)$. To this end, we need the following theorem.

\textbf{Theorem 1. } Suppose that the series
\[
S(\tau)=\mathop{\sum}\limits _{i=0}^{\infty}c_{i}e^{-d_{i}\tau},
\]
where $c_{i}\geq0$ and $d_{i}\geq0$, $i=0,1,2,\ldots$, converges
uniformly in $\tau$ and that there exist $a>0$, $b>0$, $\gamma>0$,
and $\delta>0$ such that
\[
\lim_{i\rightarrow\infty}\dfrac{c_{i}}{a^{-i}}=\gamma,\;\lim_{i\rightarrow\infty}\dfrac{d_{i}}{b^{-i}}=\delta,
\]
Then

\[
S(\tau)=\frac{\gamma}{\log b}\left(\delta\tau\right)^{-\tfrac{\log
a}{\log b}}\left(\Gamma\left(\frac{\log a}{\log
b}\right)+u\left(\log\left(\delta\tau\right)\right)\right)\left(1+o\left(1\right)\right)
\]
as $\tau\to\infty$, where

\begin{equation}
u\left(x\right)=2\mathrm{\mathrm{Re}}\mathop{\sum}\limits
_{k=1}^{+\infty}\exp\left(-i\dfrac{2\pi k}{\log
b}x\right)\Gamma\left(\frac{\log a-2\pi ik}{\log
b}\right)\label{u(x)}
\end{equation}
is a periodic function with period $\log b$ and $o\left(1\right)$
is an infinitesimal function as $\tau\rightarrow\infty$.

The proof of Theorem 1 is given in Appendix 1.

The upper bound of the function (\ref{u(x)}) is determined by the
following theorem.

\textbf{Theorem 2. } When $a>1$, $b>1$, and $a<b^{2}$, the
following inequality holds for all $x\in\mathbb{R}$:

\[
\left|u\left(x\right)\right|\leq4\pi\exp\left(-\frac{\log a}{\log b}\right)\left(\frac{\log a}{\log b}\right)^{\tfrac{\log a}{\log b}-\tfrac{1}{2}}\exp\left(\dfrac{\log b}{4\log a}\left(1+\dfrac{4\pi^{2}}{\log^{2}a}\right)^{-\tfrac{1}{2}}\right)
\]
\begin{equation}
\times\left(1+\dfrac{4\pi^{2}}{\log^{2}a}\right)^{\tfrac{\log a}{2\log b}-\tfrac{1}{4}}\dfrac{\exp\left(-\frac{2\pi}{\log b}\arctan\frac{2\pi}{\log a}\right)}{1-\exp\left(-\frac{2\pi}{\log b}\arctan\frac{2\pi}{\log a}\right)}.\label{T_2}
\end{equation}

The proof of Theorem 2 is given in Appendix 2.

Applying Theorem 1 to series (\ref{S(t)}) and setting
$c_{i}=a^{i}=p^{-i}$, $d_{i}=b^{-i}=p^{-\alpha i}$, $\gamma=1$,
$\delta=1$, and $\tau=t$, we obtain

\[
S_{\mathbb{Z}_{p}}\left(t\right)=\frac{\left(1-p^{-1}\right)}{\alpha\log p}\tau^{-\tfrac{1}{\alpha}}\left(\Gamma\left(\frac{1}{\alpha}\right)+u\left(\log t\right)\right)\left(1+o\left(1\right)\right),
\]
where
\begin{equation}
u\left(\log t\right)=2\mathrm{\mathrm{Re}}\mathop{\sum}\limits _{k=1}^{+\infty}\exp\left(-\dfrac{2\pi ik}{\alpha\log p}\left(\log t\right)\right)\Gamma\left(\frac{1}{\alpha}-\frac{2\pi ik}{\alpha\log p}\right).\label{f_logt}
\end{equation}
It is clear from the explicit expression for (\ref{f_logt}) that
this function oscillates with period $\alpha\log p$ in $\log t$
about the mean value $f=0$. Similarly, from (\ref{f(x,t)}) for
$f\left(x,t\right)$ with $\left|x\right|_{p}=p^{\gamma}$,
$\gamma>0$, we have

\[
f\left(x,t\right)=\frac{\left(1-p^{-1}\right)}{\alpha\log p}\tau^{-\tfrac{1}{\alpha}}\left(\Gamma\left(\frac{1}{\alpha}\right)+u\left(\log t-\alpha\gamma\log p\right)\right)\left(1+o\left(1\right)\right).
\]

\begin{figure}[H]
\centering{}\includegraphics[scale=0.6]{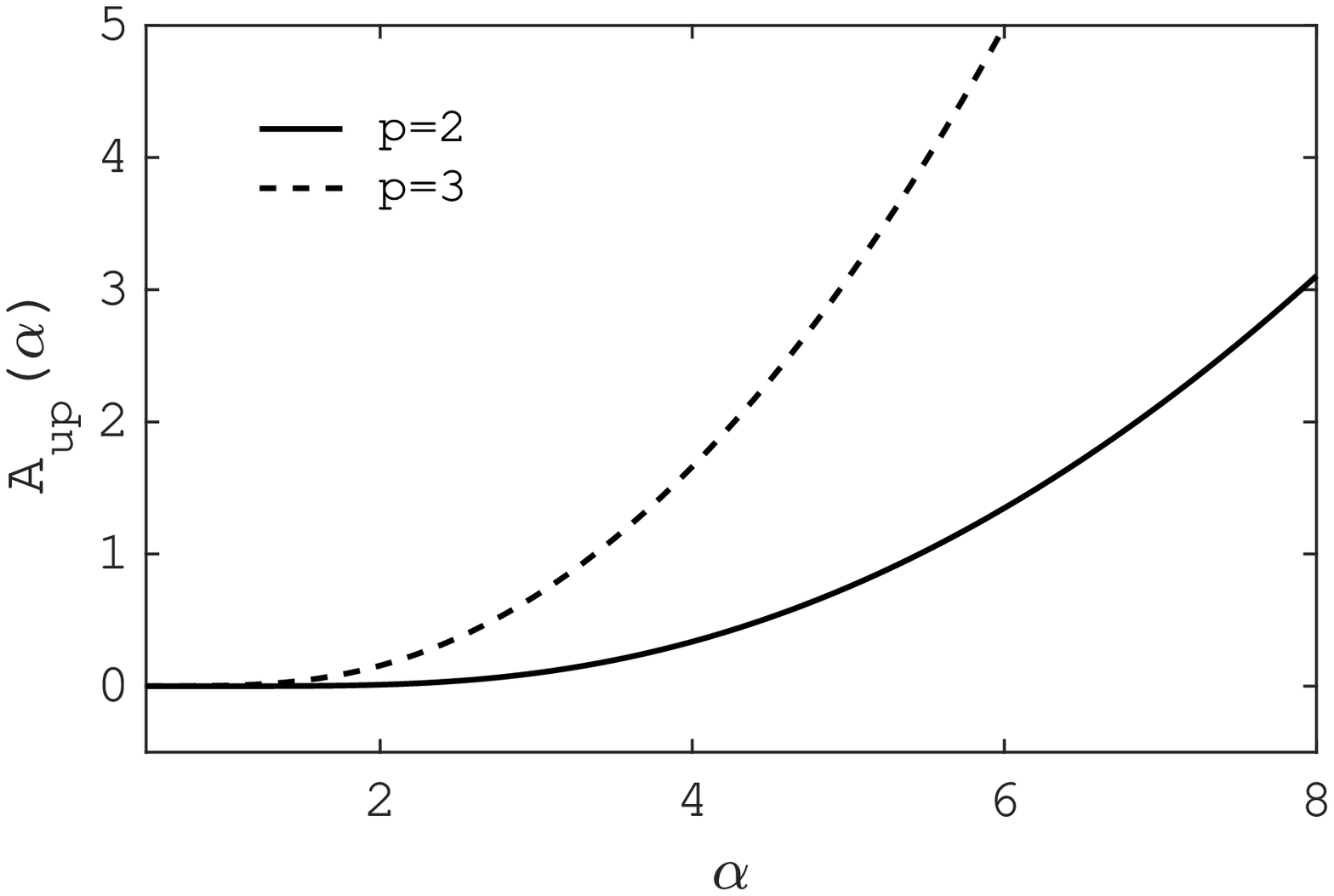}\caption{Graph of
$A_{\mathrm{up}}\left(\alpha\right)$ for $p=2$ and $p=3$.
\label{Fig_2}}
\end{figure}

Using Theorem 2, we can obtain the following upper bound for the
function (\ref{f_logt}) for $\alpha>\dfrac{1}{2}$:

\[
\left|u\left(\log t\right)\right|\leq
A_{\mathrm{up}}\left(\alpha\right),
\]
where
\[
A_{\mathrm{up}}\left(\alpha\right)=4\pi\alpha^{-\tfrac{1}{\alpha}+\tfrac{1}{2}}e^{-\tfrac{1}{\alpha}}\left(1+\frac{4\pi^{2}}{\log^{2}p}\right)^{\tfrac{1}{2\alpha}-\tfrac{1}{4}}
\]
\[
\times\exp\left(\dfrac{\alpha}{4}\left(1+\frac{4\pi^{2}}{\log^{2}p}\right)^{-\tfrac{1}{2}}\right)\dfrac{\exp\left(-\frac{2\pi}{\alpha\log p}\arctan\frac{2\pi}{\log p}\right)}{1-\exp\left(-\frac{2\pi}{\alpha\log p}\arctan\frac{2\pi}{\log p}\right)}.
\]
This formula shows that an increase in $\alpha$ leads to an
exponential growth in the upper bound of the amplitude
$\left|u\left(\log t\right)\right|$, as illustrated by the graph
of the function $A_{\mathrm{up}}\left(\alpha\right)$ for $p=2$ and
$p=3$ in Fig. \ref{Fig_2}.

\section{Oscillations in the solution of the $p$-adic diffusion equation with a reaction sink}

The solution of the equation of ultrametric diffusion with a
reaction sink with different initial conditions on a compact set
was studied in a number of works in relation to the description of
the kinetics of binding a CO molecule to myoglobin
\cite{ABKO_2002,ABZ_2014,BZ_2021}. Here we consider the simplest
version of this problem in the case when both the sink and the
initial condition are homogeneous and their support is
$\mathbb{Z}_{p}$. Thus, we consider the Cauchy problem of the form

\begin{equation}
\dfrac{df\left(x,t\right)}{dt}=-\dfrac{1}{\Gamma_{p}\left(-\alpha\right)}\intop_{\mathbb{Q}_{p}}d_{p}y\dfrac{f\left(y,t\right)-f\left(x,t\right)}{\left|x-y\right|_{p}^{\alpha+1}}-\sigma\Omega\left(\left|x\right|_{p}\right)f\left(x,t\right)\label{Vlad_s}
\end{equation}
with the initial condition (\ref{ic}), where $\sigma>0$ is the
reaction constant.

An analytical treatment of the Cauchy problem (\ref{Vlad_s}) --
(\ref{ic}) is described in detail in \cite{Bik_2010}. Here we only
consider the case of $\alpha>1$, since it is this range of values
of $\alpha$ that corresponds to the physical values of temperature
when simulating the relaxation dynamics of protein
\cite{ABZ_2014,BZ_2021}. The quantity of physical interest is
$S\left(t\right)\equiv\intop_{\mathbb{Q}_{p}}d_{p}xf\left(x,t\right)$,
which represents the probability measure of $\mathbb{Q}_{p}$. It
follows from (\ref{Vlad_s}) that the function $S\left(t\right)$ is
related to the survival probability
$S_{\mathbb{Z}_{p}}\left(t\right)\equiv\intop_{\mathbb{Z}_{p}}d_{p}xf\left(x,t\right)$
by the formula
\begin{equation}
\dfrac{dS\left(t\right)}{dt}=-\sigma S_{Z_{p}}\left(t\right).\label{d_SP}
\end{equation}

\begin{figure}[H]
\centering{}\centering{}\includegraphics[scale=0.6]{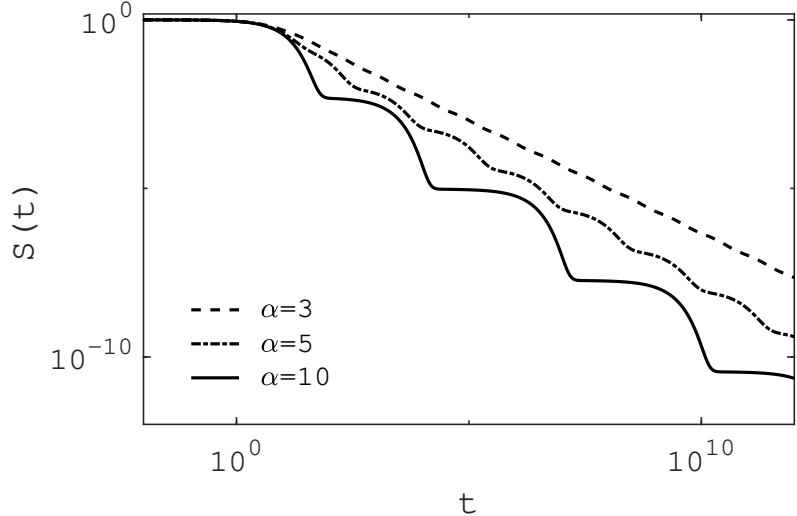}\caption{Graph
of
$S\left(t\right)\equiv\intop_{\mathbb{Q}_{p}}d_{p}xf\left(x,t\right)$,
where $f\left(x,t\right)$ is the numerical solution of equation
(\ref{Vlad_s}) on the compact set $B_{r}$ with the initial
condition (\ref{ic}) for $p=2$, $\sigma=0.1$, and $r=20$.
\label{Fig_3}}
\end{figure}

Just as in the no-sink case, the Cauchy problem (\ref{Vlad_s})
with the initial condition (\ref{ic}) admits a numerical analysis
when the integration domain $\mathbb{Q}_{p}$ is replaced by a
$p$-adic ball $B_{r}=\left\{ x:\;\left|x\right|_{p}\leq
p^{r}\right\} $. For the parameters $p=2$, $\sigma=0.1$, and
$r=20$, the graphs of the function $S\left(t\right)$ corresponding
to the values of $\alpha=3,\:5,\:10$ are demonstrated in Fig.
\ref{Fig_3} in a double logarithmic scale. We can see that the
graphs show oscillations characterized by an increase in the
amplitude and period with increasing $\alpha$.

In \cite{Bik_2010}, the first author showed that the function
$S_{Z_{p}}\left(t\right)$ can be represented as an infinite series
\begin{equation}
S_{Z_{p}}\left(t\right)={\displaystyle \sum_{k=-1}^{\infty}}b_{k}\exp\left(-\lambda_{k}t\right),\label{S_Z}
\end{equation}
where the values of $\lambda_{k}$ and $b_{k}$, $k=-1,0,1,2,...$,
are determined from the equations
\begin{equation}
1+\sigma\left.J\left(s\right)\right|_{s=-\lambda_{k}}=0,\label{eq_pol}
\end{equation}

\begin{equation}
b_{k}=\dfrac{1}{\sigma^{2}|J^{\prime}\left(-\lambda_{k}\right)|},\label{b_k}
\end{equation}
and the function $J\left(s\right)$ is given by
\[
J\left(s\right)=\left(1-p^{-1}\right){\displaystyle \sum_{k=0}^{\infty}}\dfrac{p^{-k}}{s+p^{-\alpha k}}.
\]
Using (\ref{d_SP}) and (\ref{S_Z}), we can show (see also
\cite{BZ_2021}) that
\begin{equation}
S\left(t\right)=\sigma\sum_{k=-1}^{\infty}c_{k}\exp\left(-\lambda_{k}t\right),\label{S(t)_sum}
\end{equation}
where $c_{k}=\dfrac{b_{k}}{\lambda_{k}}$.

Next, we need the following theorem.

\textbf{Theorem 3. } There exist $\gamma>0$ and $\delta>0$ such
that

\[
\lim_{k\rightarrow\infty}\dfrac{c_{k}}{p^{-\left(\alpha-1\right)k}}=\gamma,\;\lim_{i\rightarrow\infty}\dfrac{\lambda_{k}}{p^{-\alpha k}}=\delta.
\]

The proof of Theorem 3 is given in Appendix 3.

Applying Theorems 1 and 3 to (\ref{S(t)_sum}) and setting
$a=p^{\alpha-1}$, $b=p^{\alpha}$, and $\tau=t$ in Theorem 1 and
$d_{k}=\lambda_{k}$ in Theorem 3, we obtain

\[
\widetilde{S}\left(t\right)=\frac{\gamma}{\alpha\log p}\left(\delta t\right)^{-\tfrac{\alpha-1}{\alpha}}\left(\Gamma\left(\frac{\alpha-1}{\alpha}\right)+u\left(\log\left(\delta t\right)\right)\right)\left(1+o\left(1\right)\right),
\]
where

\[
u\left(\log\left(\delta t\right)\right)=2\mathrm{\mathrm{Re}}\mathop{\sum}\limits _{k=1}^{+\infty}\exp\left(-i\dfrac{2\pi k}{\alpha\log p}\log\left(\delta t\right)\right)\Gamma\left(\frac{\left(\alpha-1\right)\log p-2\pi ik}{\alpha\log p}\right)
\]
is a function periodic in $\log\left(\delta t\right)$ with period
$\alpha\log p$.

\section{Discussion}

We have considered the solution of the Cauchy problem with the
initial condition on a compact set in two cases: for purely
$p$-adic diffusion and for $p$-adic diffusion with a reaction
sink. We have shown that the large-time solution in these two
cases is described by a power law on which logarithmic
oscillations are superimposed. We emphasize that the very presence
of oscillations in the solution does not mean its periodicity; it
only means a periodic deviation of this solution from the main
trend described by a decreasing power-law function.

As already mentioned, the presence of oscillations in ultrametric
models describing the relaxation dynamics of protein is associated
with the hierarchy of the space of conformational states of
protein. From physical considerations, the emergence of these
oscillations in the hierarchical model of pure random walk of
protein on conformations can be explained as follows. Suppose that
at the initial time protein is in some basin $B_{0}$ of
conformational states and that, in the space of all conformational
states, one can select basins $B_{1}$, $B_{2}$, $\ldots$ that are
hierarchically embedded in each other: $B_{0}\subset B_{1}\subset
B_{2}\subset\ldots$. Suppose also that the basins $B_{0}$,
$B_{1}$, $B_{2}$, $\ldots$ are separated from each other by energy
barriers $E_{1}$, $E_{2}$, $E_{3}$, $\ldots$ and that $E_{1}\ll
E_{2}\ll E_{3}\ll\ldots$. At the initial time, the support of the
distribution function coincides with $B_{0}$, and the probability
measure ``spreads'' over the whole conformation space in the
course of time. The average time to overcome the energy barrier
$E_{i}$ separating the $i$th basin from the basins embedded in it
is proportional to the Boltzmann factor
$\tau_{i}\sim\exp\left(\alpha E_{i}\right)$, where
$\alpha=\dfrac{1}{kT}$, $k$ is the Boltzmann constant, and $T$ is
temperature; in this case it is obvious that
$\tau_{1}\ll\tau_{2}\ll\tau_{3}\ll\ldots$.

Consider a basin $B_{i-1}$ and denote by $S_{i-1}\left(t\right)$
the probability measure of this basin. Suppose that the function
$S_{i-1}\left(t\right)$ can be approximated by a simple function
$S_{i-1,\mathrm{trend}}\left(t\right)$ (for example, by a
decreasing power trend) with a sufficiently high degree of
accuracy. Consider a time instant $t$ such that $\tau_{i-1}\ll
t\ll\tau_{i}$. Since $t\ll\tau_{i}$, the probability measure at
this time instant $t$ is mainly concentrated in the basin
$B_{i-1}$ because the average time to overcome the energy barrier
separating the basin $B_{i-1}$ from the set $B_{i}\setminus
B_{i-1}$ is much greater than $t$. Moreover, since
$t\gg\tau_{i-1}$, we can assume that the distribution in $B_{i-1}$
at these instants of time is close to the quasi-equilibrium
distribution because the probability of transition from $B_{i-1}$
to $B_{i}\setminus B_{i-1}B_{i-1}$ is small compared to the
transitions inside $B_{i-1}$.

Further, as time proceeds from $\tau_{i}$ to $\tau_{i+1}$, the
region in which the the main part of the probability measure is
concentrated increases, and, due to the conservation of
probability measure in the whole space of conformational states,
the distribution function increases in the set $B_{i}\setminus
B_{i-1}$ and decreases in $B_{i-1}$. Then at times $t$ such that
$\tau_{i}\ll t\ll\tau_{i+1}$, the probability measure is mainly
concentrated in the basin $B_{i}$, and this distribution in
$B_{i}$ is also close to the quasi-equilibrium distribution. Since
the variation rate of the distribution function decreases as it
approaches the equilibrium state, it is natural to assume that,
for all $j>i$ at time instants $t$ such that $\tau_{j-1}\ll
t\ll\tau_{j}$, the decrease rate of the function
$S_{i-1}\left(t\right)$ with respect to the decrease rate of the
trend $S_{i-1,\mathrm{trend}}\left(t\right)$ should be less than
its decrease rate with respect to the decrease rate of the trend
at time instants $t$ satisfying the conditions
$\tau_{j-1}<t\ll\tau_{j}$ or $\tau_{j-1}\ll t<\tau_{j}$. It is
this change in the decrease rate of the function
$S_{i-1}\left(t\right)$ with respect to the function
$S_{i-1,\mathrm{trend}}\left(t\right)$ that manifests itself as a
superimposition of oscillations on the trend
$S_{i-1,\mathrm{trend}}\left(t\right)$. It is obvious that the
period $\tau_{i}-\tau_{i-1}$ of these oscillations increases with
time. In the general case of arbitrary relation between the times
$\tau_{i}$ (which corresponds to an arbitrary relation between the
energy barriers $E_{i}$), these oscillations are aperiodic.
Nevertheless, if the times $\tau_{i}$ are related by
$\tau_{i}=\tau_{0}\theta^{i}$, where $\tau_{0}$ and $\theta$ are
some parameters, then the period
$\lambda_{i}-\lambda_{i-1}=\log\theta$ as a function of the
variable $\lambda=\log t$ does not depend on time. This means that
the decrease rate of the distribution function oscillates with a
period, which increases in time by a logarithmic law. Hence we can
also conclude that, as temperature decreases, the logarithmic
oscillations become more pronounced. Indeed, as temperature
decreases, the ratio $\dfrac{E_{i}}{E_{i-1}}$, and hence also
$\dfrac{\tau_{i}}{\tau_{i-1}}$ increase. Hence, the lower the
temperature, the closer to equilibrium the state of protein is in
the region $B_{i-1}$ at time instants from the interval close to
$\tau_{i}$ $\left(\tau_{i-1},\tau_{i}\right)$. However, the closer
the state of protein to equilibrium, the lower the decrease rate
of the distribution function in any region contained in the basin
$B_{i-1}$. Thus, at sufficiently low temperatures, one should
observe regions of decreasing relaxation curves, which are close
to horizontal lines in the logarithmic scale (see, for example,
the curves in Figs. \ref{Fig_1} and \ref{Fig_3} corresponding to
$\alpha=10$).

Naturally, the observation of logarithmic oscillations on
experimental relaxation curves can provide a convincing support
for the assumption of ultrametricity of the space of
conformational states of protein and the adequacy of $p$-adic
models for the description of relaxation dynamics of protein. In
this context we emphasize once again that the $p$-adic diffusion
equation (\ref{Vlad}) is fundamental for the description of the
conformational dynamics of protein and underlies the description
of experiments on spectral diffusion
\cite{AB_2008,ABZ_2014,BZ_2022}, while the $p$-adic diffusion
equation with a reaction sink (\ref{Vlad_s}) underlies the
description of experiments on the kinetics of $\mathrm{CO}$
binding to myoglobin \cite{ABKO_2002,ABZ_2013,ABZ_2014,BZ_2021}.
Hence, the presence of logarithmic oscillations in the solutions
of these equations should also manifest itself to some extent in
the relaxation curves of experiments that can be described within
the models of $p$-adic diffusion.

An analysis of experimental studies on the conformational dynamics
of protein shows that there seem to be no experiments directly
aimed at finding the oscillations of relaxation curves. The study
of the graphs presented in a number of experimental works in
principle allows one to carry out an approximate analysis of the
presence of oscillations within the available accuracy. As pointed
out in \cite{MKJ}, it is convenient to consider instead of the
relaxation function $S\left(t\right)$ the logarithmic derivative
of its logarithm, i.e., the function $\dfrac{d\log
S\left(t\right)}{d\log t}$, which should demonstrate purely
logarithmic oscillations in the neighborhood of its mean value.
Nevertheless, a similar analysis on the detection of logarithmic
oscillations in the behavior of relaxation curves with a power-law
trend presents a problem, first, due to the small amplitude of
these oscillations and, second, due to the presence of rather
large errors in the location of experimental points. As an
example, below we present an attempt to detect logarithmic
oscillations in some experimental relaxation curves obtained in
\cite{LJA_1993,JLA_1994,Sabelko_1,Sabelko_2}.

\begin{figure}[H]
\centering{}\includegraphics[scale=0.5]{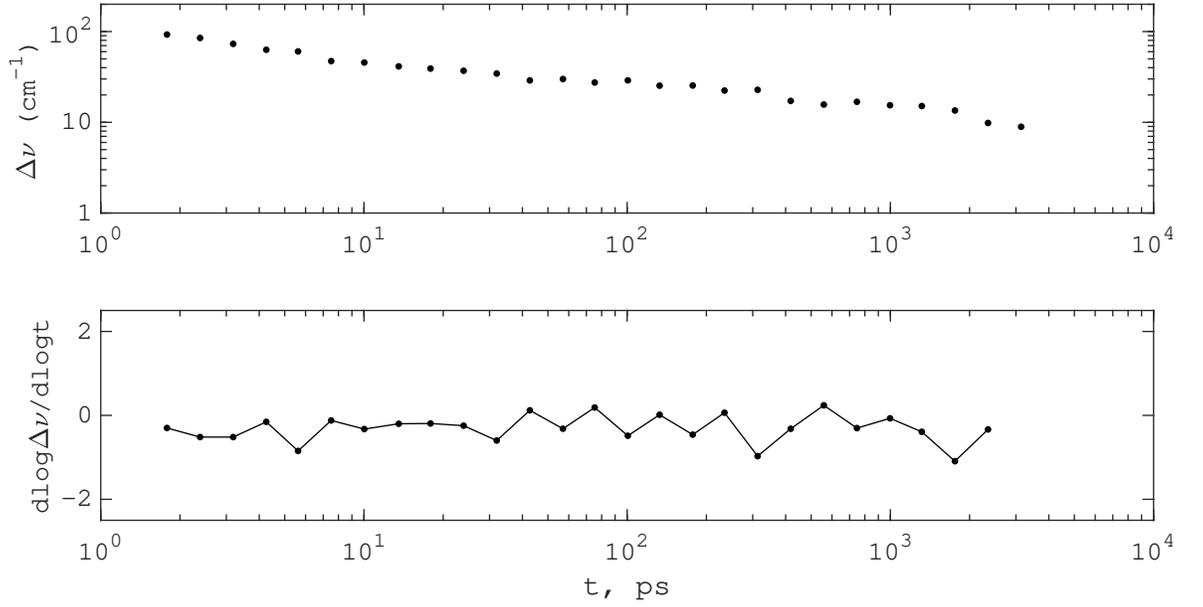}\caption{Top:
experimental dependence of the variation $\Delta\nu\left(t\right)$
in the position of band III after photodissociation
$\mathrm{MbCO}$, shown in Fig. 3 in \cite{LJA_1993}; bottom: the
function $\dfrac{d\log\Delta\nu\left(t\right)}{d\log t}$
constructed using the function $\Delta\nu\left(t\right)$.
\label{Fig_4}}
\end{figure}

In \cite{LJA_1993,JLA_1994}, the authors investigated the
evolution of the absorption spectrum of the heme group of
myoglobin ($\mathrm{Mb}$) in the tertiary structure at room
temperature after photodissociation
$\mathrm{MbCO}\rightarrow\mathrm{Mb+CO}$. The observations were
made on the so-called band III, which is responsible for the
porphyrin-$\pi$ $\rightarrow$ iron-d charge-transfer transition
with a frequency of about $13110\:\mathrm{cm}^{-1}$
($763\:\mathrm{nm}$). The variation of conformational states was
identified with the variation $\Delta\nu$ in the position of band
III relative to its position in the equilibrium state
$\mathrm{Mb}$. In \cite{LJA_1993,JLA_1994}, the authors presented
experimental graphs of the function $\Delta\nu\left(t\right)$ for
different lengths of the laser pulse and in different observation
time windows. In \cite{Frau_M}, using the data for
$\Delta\nu\left(t\right)$ obtained in \cite{JLA_1994}, the authors
illustrated the presence of oscillations in the function
$\dfrac{d\log\Delta\nu\left(t\right)}{d\log t}$, which, according
to the authors, indicate the hierarchy of protein relaxation. Here
we present a similar analysis carried out for the experimental
data obtained in \cite{LJA_1993}. Figure \ref{Fig_4} (top) shows
the graph of the experimentally measured dependence of the
variation $\Delta\nu$ in the position of band III on time, which
was presented in \cite{JLA_1994}. Figure 4 (bottom) demonstrates
the function $\dfrac{d\log\Delta\nu\left(t\right)}{d\log t}$,
which exhibits an oscillation-like behavior.

\begin{figure}[H]
\centering{}\includegraphics[scale=0.5]{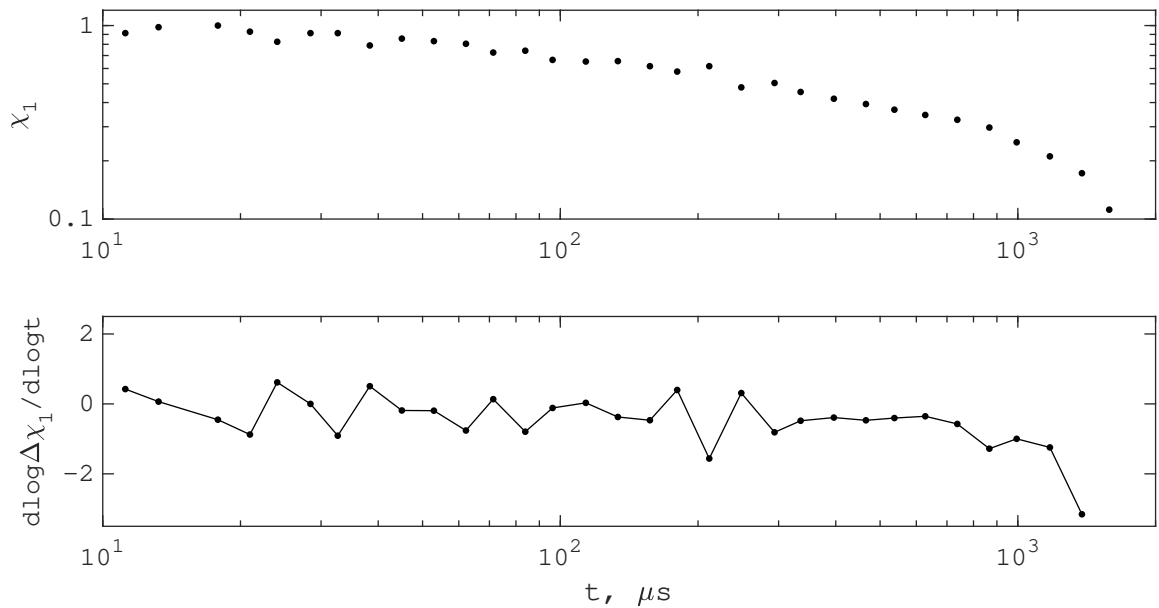}\caption{Top:
experimental function $\chi_{1}\left(t\right)$ for
$\mathrm{hisPGK}$ protein, presented in Fig. 6 in \cite{Sabelko_2}
for $T_{1}=-12\,^{\circ}\mathrm{C}$ and
$T_{2}=+3\,^{\circ}\mathrm{C}$; bottom: the function
$\dfrac{d\log\chi_{1}\left(t\right)}{d\log t}$ constructed using
the function $\chi_{1}\left(t\right)$. \label{Fig_5}}
\end{figure}

The emergence of logarithmic oscillations was also mentioned in
\cite{Sabelko_1,Sabelko_2}, where the authors presented the
results of experiments on the folding kinetics of phosphoglycerate
kinase protein ($\mathrm{PGK}$). In these experiments, a protein
sample was subjected to cold denaturation in supercooled aqueous
buffer at temperature $T_{1}$. Then the aqueous buffer was rapidly
heated by an infrared pulse to temperature $T_{2}$, after which
the protein in the heated buffer started to fold. During folding
in the time window $t\in\left(0,\:1000\:\mu s\right)$ in the
experiment, in each time interval $\tau=14\:\mathrm{ns}$ protein
was irradiated by a nanosecond laser pulse, which led to the
fluorescence of protein. The time profile of this fluorescence was
described by a function $f\left(t^{\prime}\right)$, which rapidly
decreased in time $t^{\prime}\in\left(0,\tau\right)$. The state of
the protein at time $t$ was determined by the function
$f\left(t,t^{\prime}\right)$, which was approximated as
$f\left(t,t^{\prime}\right)=C_{1}\left(t\right)f_{1}\left(t^{\prime}\right)+C_{2}\left(t_{i}\right)_{1}f_{2}\left(t^{\prime}\right)$,
where $t^{\prime}\in\left(0,\tau\right)$; the function
$f_{1}\left(t\right)$ described the fluorescence profile in the
initial (denaturated) state, and $f_{2}\left(t\right)$, in the
final (folded) state. The observable parameter $\chi_{1}$ shown in
the graphs in these studies represented the ratio
$\chi_{1}\left(t\right)=\dfrac{C_{1}\left(t\right)}{C_{1}\left(t\right)+C_{2}\left(t\right)}$.
Figure \ref{Fig_5} (top) shows the graph of of the normalized
quantity $\chi_{1}\left(t\right)$ ($\max\chi_{1}=1$), which is
plotted using the data presented in Fig. 6 in \cite{Sabelko_2} for
$\mathrm{hisPGK}$ protein. Figure \ref{Fig_5} (bottom) presents
the function $\dfrac{d\log\chi_{1}\left(t\right)}{d\log t}$, which
also exhibits an oscillation-like behavior.

Naturally, the analysis carried out above of relaxation curves in
two different experiments can in no way serve as a proof of the
presence of logarithmic oscillations in these processes. First of
all, note that the experimental data presented by dots in the
graphs have a large dispersion due to experimental errors. For
example, one should note the local growth of relaxation curves in
separate regions in the upper graphs. Obviously, this is an
artifact of the measurement process and should not occur in
reality. Therefore, the oscillations shown in the lower graphs in
Figs. \ref{Fig_4} and \ref{Fig_5} are strongly distorted, even if
they have to do with real oscillations.

Summarizing, we can state that the currently available
experimental data on the relaxation dynamics of protein do not
allow one to make an unambiguous conclusion about the presence of
logarithmic oscillations in the behavior of relaxation curves.
Thus, the problem of experimental verification of the presence (or
absence) of oscillations in the evolution of conformational
rearrangements of a protein molecule still remains topical, since
its solution would provide a deeper insight into the applicability
of the ultrametric concept to the description of the
conformational dynamics of protein.

\section*{Acknowledgments}

The study was supported in part by the Ministry of Education and Science
of Russia by State assignment to educational and research institutions
under Projects No. FSSS-2023-0009

\section*{Data Availability Statement}

The data supporting the findings of this study are available within
the article and its supplementary material. All other relevant source
data are available from the corresponding author upon reasonable request.

\section*{Appendix 1}

In this appendix, we present the proof of Theorem 1. Consider the
series

\noindent
\begin{equation}
F(\tau)=\gamma\mathop{\sum}\limits _{i=0}^{\infty}a^{-i}e^{-\delta b^{-i}\tau},\label{A1}
\end{equation}
.

\noindent which obviously converges uniformly in $\tau$. Then
$\mathop{\sum}\limits
_{i=0}^{\infty}\left(a^{-i}e^{-b^{-i}\tau}-c_{i}e^{-d_{i}\tau}\right)$
also converges uniformly in $\tau$ and
\[
\lim_{\tau\rightarrow\infty}\left(F(\tau)-S(\tau)\right)=\mathop{\sum}\limits
_{i=0}^{\infty}\lim_{\tau\rightarrow\infty}\left(\gamma
a^{-i}e^{-\delta b^{-i}\tau}-c_{i}e^{-d_{i}\tau}\right)=0;
\]
hence it follows that
\begin{equation}
S(\tau)=F(\tau)\left(1+o\left(1\right)\right)\label{A_2}
\end{equation}
as $\tau\to\infty$, where $o\left(1\right)$ is an infinitesimal
function as $\tau\rightarrow\infty$. Rewriting (\ref{A1}) as
$F(\tau)=\mathop{\sum}\limits _{m=0}^{\infty}g(m),$ where
$g(m)=\gamma a^{-m}e^{-\delta b^{-m}\tau}$, we apply the
well-known Poisson summation formula

\noindent
\[
\mathop{\sum}\limits _{m=0}^{\infty}g(m)=\mathop{\sum}\limits _{k=-\infty}^{+\infty}\mathop{\smallint}\limits _{0}^{\infty}g(x)\exp(2\pi ikx)dx.
\]
Then

\noindent
\begin{equation}
F(\tau)=\gamma\mathop{\sum}\limits _{k=-\infty}^{+\infty}\mathop{\smallint}\limits _{0}^{\infty}a^{-x}\exp\left(-\delta b^{-x}\tau+2\pi ikx\right)dx.\label{A3}
\end{equation}

\noindent Let us change the variable $y=\delta b^{-x}\tau$ in the
integral in (\ref{A3}). After straightforward transformations we
arrive at the expression
\begin{equation}
F(\tau)=\frac{\gamma}{\log b}\tau^{-\tfrac{\log a}{\log b}}\mathop{\sum}\limits _{k=-\infty}^{+\infty}\left(\delta\tau\right)^{\tfrac{-2\pi ik}{\log b}}\gamma\left(\frac{\log a-2\pi ik}{\log b},\,\delta\tau\right),\label{A4}
\end{equation}

\noindent where $\gamma(z,\,t)=\intop_{0}^{t}y^{z-1}e^{-y}dy$ is
the incomplete gamma function. As $\tau\rightarrow\infty$,
\[
\gamma\left(\frac{\log a-2\pi ik}{\log b},\,\delta\tau\right)=\Gamma\left(\frac{\log a-2\pi ik}{\log b}\right)\left(1+o\left(1\right)\right),
\]
where $\Gamma\left(z\right)=\intop_{0}^{\infty}t^{z-1}e^{-t}dt$ is
the gamma function, and we have

\noindent
\begin{equation}
F(\tau)=\frac{\gamma}{\log b}\left(\delta\tau\right)^{-\tfrac{\log a}{\log b}}\left(\Gamma\left(\frac{\log a}{\log b}\right)+2\mathrm{\mathrm{Re}}\mathop{\sum}\limits _{k=1}^{+\infty}\left(\delta\tau\right)^{\tfrac{-2\pi ik}{\log b}}\Gamma\left(\frac{\log a-2\pi ik}{\log b}\right)\right)\left(1+o\left(1\right)\right),\label{A5}
\end{equation}
which, combined with (\ref{A_2}), proves the theorem.

\section*{Appendix 2}

In this appendix, we present the proof of Theorem 2. It follows
from the explicit form of the function (\ref{u(x)})
\[
u\left(x\right)=2\mathrm{\mathrm{Re}}\mathop{\sum}\limits _{k=1}^{+\infty}\exp\left(-i\dfrac{2\pi k}{\log b}x\right)\Gamma\left(\frac{\log a-2\pi ik}{\log b}\right)
\]
that
\begin{equation}
\left|u\left(x\right)\right|\leq2\mathop{\sum}\limits
_{k=1}^{+\infty}\left|\Gamma\left(z\right)\right|,\label{B_0}
\end{equation}
where $z=\dfrac{\log a}{\log b}-\dfrac{2\pi ik}{\log b}$; in this
case, $\arg z=-\arctan\dfrac{2\pi k}{\log a}$,
$-\dfrac{\pi}{2}<\arg z<0$, and $\left|z\right|=\dfrac{\log
a}{\log
b}\left(1+\dfrac{4\pi^{2}k^{2}}{\log^{2}a}\right)^{\tfrac{1}{2}}$.
Let us apply Stirling's formula
\begin{equation}
\Gamma\left(z\right)=2\pi z^{z-\tfrac{1}{2}}e^{-z}\exp\left(\mu\left(z\right)\right)\;\mathrm{for}\;\left|\arg z\right|<\pi,\label{S_S}
\end{equation}
where $\mu\left(z\right)$ has the Stieltjes representation
\[
\mu\left(z\right)=\intop_{0}^{\infty}\dfrac{Q\left(t\right)}{\left(z+t\right)^{2}}dt,\;Q\left(t\right)=\dfrac{1}{2}\left(t-\left[t\right]-\left(t-\left[t\right]\right)^{2}\right),
\]
and $\left[t\right]$ denotes the greatest integer $<t$ (see, for
example, \cite{Remmert}). Taking into account the inequality
$0\leq Q\left(t\right)\leq\dfrac{1}{8}$, we can obtain the
following upper bound for the function

\[
\left|\exp\left(\mu\left(z\right)\right)\right|=\exp\left(\intop_{0}^{\infty}\dfrac{Q\left(t\right)}{\left|z+t\right|^{2}}dt\right)\leq\exp\left(\dfrac{1}{8}\intop_{0}^{\infty}\dfrac{dt}{\left|z+t\right|^{2}}\right).
\]
Since
\[
\left|z+t\right|^{2}=\left|z\right|^{2}+2\left|z\right|\cos\left(\arg z\right)t+t^{2}=
\]

\[
=\left(\left|z\right|+t\right)^{2}-4\left|z\right|t\left(1-\cos^{2}\left(\dfrac{\arg z}{2}\right)\right)
\]
and since $\left(\left|z\right|+t\right)^{2}\geq4\left|z\right|t$,
it follows that

\[
\left|z+t\right|^{2}\geq\left(\left|z\right|+t\right)^{2}\cos^{2}\left(\dfrac{\arg z}{2}\right).
\]
Then

\[
\left|\exp\left(\mu\left(z\right)\right)\right|\leq\exp\left(\dfrac{1}{8\cos^{2}\left(\dfrac{\arg z}{2}\right)}\intop_{0}^{\infty}\dfrac{dt}{\left(\left|z\right|+t\right)^{2}}\right)=\exp\left(\dfrac{1}{8\left|z\right|\cos^{2}\left(\dfrac{\arg z}{2}\right)}\right).
\]
Since $\cos\arg z>0$, it follows that $\cos^{2}\left(\dfrac{\arg
z}{2}\right)=\dfrac{1}{2}+\dfrac{1}{2}\cos\arg z\geq\dfrac{1}{2}$
and we have

\begin{equation}
\left|\exp\left(\mu\left(z\right)\right)\right|\leq\exp\left(\dfrac{1}{4\left|z\right|}\right)=\exp\left(\dfrac{\log b}{8\log a}g_{k}^{-\tfrac{1}{2}}\right),\label{B_1}
\end{equation}
where $g_{k}=1+\dfrac{4\pi^{2}k^{2}}{\log^{2}a}$. In view of
(\ref{S_S}),

\[
\left|\Gamma\left(z\right)\right|=2\pi\left|e^{z\log z}\right|\left|z\right|^{-\tfrac{1}{2}}\left|e^{-z}\right|\left|\exp\left(\mu\left(z\right)\right)\right|\leq
\]
\[
\leq2\pi\left|\exp\left(\left(\frac{\log a}{\log b}-\frac{2\pi ik}{\log b}\right)\left(\log\left(\frac{\log a}{\log b}g_{k}^{\tfrac{1}{2}}\right)-i\arctan\frac{2\pi k}{\log a}\right)\right)\right|
\]
\[
\times\left(\frac{\log b}{\log a}\right)^{\tfrac{1}{2}}g_{k}^{-\tfrac{1}{4}}\exp\left(-\frac{\log a}{\log b}\right)\exp\left(\dfrac{\log b}{4\log a}g_{k}^{-\tfrac{1}{2}}\right)
\]

\begin{equation}
\leq2\pi\exp\left(-\frac{\log a}{\log b}\right)\left(\frac{\log a}{\log b}\right)^{\tfrac{\log a}{\log b}-\tfrac{1}{2}}g_{k}^{\tfrac{\log a}{2\log b}-\tfrac{1}{4}}\exp\left(-h_{k}\right)\exp\left(\dfrac{\log b}{4\log a}g_{k}^{-\tfrac{1}{2}}\right),\label{B_2}
\end{equation}
where $h_{k}=\dfrac{2\pi k}{\log b}\arctan\dfrac{2\pi k}{\log a}$.
Then from (\ref{B_0}) we have
\[
\left|u\left(x\right)\right|\leq4\pi\exp\left(-\frac{\log a}{\log b}\right)\left(\frac{\log a}{\log b}\right)^{\tfrac{\log a}{\log b}-\tfrac{1}{2}}
\]
\[
\times\exp\left(\dfrac{\log b}{4\log a}g_{k}^{-\tfrac{1}{2}}\right)g_{k}^{\tfrac{\log a}{2\log b}-\tfrac{1}{4}}\mathop{\sum}\limits _{k=1}^{+\infty}\exp\left(-\dfrac{2\pi k}{\log b}\arctan\dfrac{2\pi}{\log a}\right),
\]
which implies the assertion of Theorem 2.

\section*{Appendix 3}

In this appendix, we present the proof of Theorem 3.

Taking into account that ${\displaystyle
\lim_{\sigma\rightarrow0}}\lambda_{k}=p^{-\alpha\left(k+1\right)}$
(see the solution of the Cauchy problem (\ref{Vlad}) -- (\ref{ic})
in the absence of a sink), it is convenient to use the following
representation for $\lambda_{k}$:

\begin{equation}
\lambda_{k}=p^{-\alpha k}\left(p^{-\alpha}+\Delta_{k}\right).\label{lambda_k_delta_k}
\end{equation}
It can be shown from the graphical analysis of the equation
$1+\sigma J\left(s\right)=0$ that $0<\Delta_{k}<1-p^{-\alpha}$.
Then (\ref{lambda_k_delta_k}) implies the following estimate for
$\lambda_{k}$: $p^{-\alpha k}p^{-\alpha}<\lambda_{k}<p^{-\alpha
k}$. From the equation $1+\sigma J\left(-\lambda_{k}\right)=0$ we
have

\begin{equation}
{\displaystyle \sum_{n=0}^{\infty}}\dfrac{p^{-n}}{p^{-\alpha n}-\lambda_{k}}=-\dfrac{p}{p-1}\dfrac{1}{\sigma}.\label{eq_lambda_k}
\end{equation}
Singling out the ($k+1$)th term in the sum on the left-hand side
of equation (\ref{eq_lambda_k}), multiplying it by $p^{k}$ and
dividing by $p^{\alpha k}$, and taking into account
(\ref{lambda_k_delta_k}), we can rewrite (\ref{eq_lambda_k}) as
\begin{equation}
\dfrac{p^{-1}}{\Delta_{k}}=\dfrac{p}{p-1}\dfrac{p^{-\left(\alpha-1\right)k}}{\sigma}+\sum_{i=0}^{k}\dfrac{p^{i}}{p^{\alpha i}-p^{-\alpha}-\Delta_{k}}{\displaystyle -\sum_{j=2}^{\infty}\dfrac{p^{-j}}{p^{-\alpha}-p^{-\alpha j}+\Delta_{k}}}.\label{eq_poles}
\end{equation}

For further proof we need the following two lemmas.

\textbf{Lemma 1.}

\begin{equation}
\dfrac{1-p^{-\alpha+1}}{1+p}\left(1-p^{-\alpha}\right)<\Delta_{k}<\dfrac{1-p^{-\alpha}}{p}.\label{App_A_Result}
\end{equation}

\textbf{Proof. } The upper bound for $\Delta_{k}$ follows from the
inequality

\begin{equation}
\dfrac{p^{-1}}{\Delta_{k}}\text{>}{\displaystyle \dfrac{1}{1-p^{-\alpha}-\varDelta_{k}}{\displaystyle -\dfrac{p^{-1}}{p-1}\dfrac{1}{\varDelta_{k}}}},\label{B_neq_1}
\end{equation}
which, in turn, follows from (\ref{eq_poles}). The lower bound for
$\Delta_{k}$ follows from the inequality

\begin{equation}
\dfrac{p^{-1}}{\Delta_{k}}<\dfrac{p}{p-1}\dfrac{p^{-\left(\alpha-1\right)k}}{\sigma}+{\displaystyle {\displaystyle \dfrac{1}{1-p^{\alpha}-\Delta_{k}}\dfrac{p^{\left(\alpha-1\right)}}{p^{\left(\alpha-1\right)}-1}},}\label{B_neq_2}
\end{equation}
which also follows from (\ref{eq_poles}). Lemma 1 is proved.

\textbf{Lemma 2.} There exists the limit
$\lim_{k\rightarrow\infty}\Delta_{k}\equiv\Delta>0$.

\textbf{Proof. } By Lemma 1, a sequence
$\dfrac{p^{-1}}{\Delta_{k}}$ is defined that is also bounded.
Suppose that the limit $\lim_{k\rightarrow\infty}\Delta_{k}$ does
not exist. Then the limit of the sequence
$\dfrac{p^{-1}}{\Delta_{k}}$ as $k\rightarrow\infty$ also does not
exist. In view of the statement logically opposite to the Cauchy
criterion, this means that there exists a number $\varepsilon>0$
such that, for any $K\in\mathbb{Z}_{+}$, there exist
$k,\:n\in\mathbb{Z}_{+}$, $k>K$, such that
$\left|\dfrac{p^{-1}}{\Delta_{k}}-\dfrac{p^{-1}}{\Delta_{k+n}}\right|>\varepsilon$
(or
$\left|\Delta_{k+n}-\Delta_{k}\right|>\varepsilon'=p\dfrac{\left(1-p^{-\alpha+1}\right)^{2}\left(1-p^{-\alpha}\right)^{2}}{\left(1+p\right)^{2}}\varepsilon$).

Consider the difference

\[
\dfrac{p^{-1}}{\Delta_{k}}-\dfrac{p^{-1}}{\Delta_{k+n}}=\dfrac{p}{p-1}\dfrac{p^{-\left(\alpha-1\right)k}}{\sigma}\left(1-p^{-\left(\alpha-1\right)n}\right)
\]

\[
-\sum_{i=1}^{n}\dfrac{p^{k}p^{i}}{p^{\alpha k}p^{\alpha i}-p^{-\alpha}-\Delta_{k+n}}
\]
\[
+\sum_{i=0}^{k}\left(\dfrac{p^{i}}{p^{\alpha i}-p^{-\alpha}-\Delta_{k}}{\displaystyle -\dfrac{p^{i}}{p^{\alpha i}-p^{-\alpha}-\Delta_{k+n}}}\right)
\]
\begin{equation}
+\sum_{j=2}^{\infty}\left(\dfrac{p^{-j}}{p^{-\alpha}-p^{-\alpha j}+\Delta_{k+n}}-\dfrac{p^{-j}}{p^{-\alpha}-p^{-\alpha j}+\Delta_{k}}\right).\label{del_del}
\end{equation}
Irrespective of $\Delta_{k}$ and $\Delta_{k+n}$, for sufficiently
large $K$ the terms on the right-hand side of equality
(\ref{del_del}) in the first and second rows can be made
arbitrarily small. Conversely, the terms in the third and fourth
rows in (\ref{del_del}) depend on the relation between
$\Delta_{k}$ and $\Delta_{k+n}$. For
$\left|\Delta_{k+n}-\Delta_{k}\right|>\varepsilon'$, the absolute
values of these terms cannot be made arbitrarily small for an
indefinite increase in $K$, and they are bounded from below.
Moreover, notice that the sign of these terms in the third and
fourth rows of expression (\ref{del_del}) is opposite to the sign
of the difference
$\dfrac{p^{-1}}{\Delta_{k}}-\dfrac{p^{-1}}{\Delta_{k+n}}$. This
means that, for sufficiently large $K$, there always exist
$k,\:n\in\mathbb{Z}_{+}$, $k>K$, such that the right- and
left-hand sides of (\ref{del_del}) have opposite signs, which
leads to a contradiction. This contradiction proves Lemma 2.

It follows from Lemma 2 that there exists the limit
\[
\lim_{i\rightarrow\infty}\dfrac{\lambda_{k}}{p^{-\alpha k}}=\lim_{i\rightarrow\infty}\dfrac{p^{-\alpha k}\left(p^{-\alpha}+\Delta_{k}\right)}{p^{-\alpha k}}=p^{-\alpha}+\Delta\equiv\delta.
\]
Next, for $b_{k}$ we can write

\[
b_{k}=\dfrac{\sigma^{-2}}{P\left(\lambda_{k}\right)},
\]
where

\[
P\left(\lambda_{k},\alpha\right)=\dfrac{dJ\left(s\right)}{ds}|_{s=-\lambda_{k}}=\left(1-p^{-1}\right)\sum_{n=0}^{\infty}\dfrac{p^{-n}}{\left(p^{-\alpha n}-\lambda_{k}\right)^{2}}=
\]
\[
=\left(p-1\right)\dfrac{p^{\left(2\alpha-1\right)k}}{\Delta_{k}^{2}}\left(1+o\left(1\right)\right),
\]
and where $o\left(1\right)$ is an infinitesimal sequence as
$k\rightarrow\infty$. Hence,
\[
b_{k}=\dfrac{\sigma^{-2}\Delta_{k}^{2}}{p-1}p^{-\left(2\alpha-1\right)k}\left(1+o\left(1\right)\right),
\]
\[
c_{k}=\dfrac{b_{k}}{\lambda_{k}}=\dfrac{\sigma^{-2}\Delta_{k}^{2}}{\left(p^{-\alpha}+\Delta_{k}\right)\left(p-1\right)}p^{-\left(\alpha-1\right)k}\left(1+o\left(1\right)\right)
\]
and thus
\[
\lim_{k\rightarrow\infty}\dfrac{c_{k}}{p^{-\left(\alpha-1\right)k}}=\dfrac{\sigma^{-2}\Delta^{2}}{\delta\left(p-1\right)}\equiv\gamma,
\]
which proves Theorem 3.


\begin{thebibliography}{10}
\bibitem{Blatz} A. L. Blatz and K.L. Magleby, ``Quantitative description
of three modes of activity of fast chloride channels from rat
skeletal muscle,'' The Journal of Physiology \textbf{378} (1),
141--174 (1986).

\bibitem{Anifrani} J. C. Anifrani, C. Le Floc'h, D. Sornette and
B. Souillard, ``Universal Log-periodic correction to renormalization
group scaling for rupture stress prediction from acoustic emissions,''
J. Phys. I France \textbf{5}, 631--638 (1995).

\bibitem{Sahimi} M. Sahimi and S. Arbabi, ``Scaling laws for fracture
of heterogeneous materials and rocks,'' Phys. Rev. Lett. \textbf{77},
3689--3692 (1996).

\bibitem{Johansen_1} A. Johansen, H. Saleur and D. Sornette, ``New
evidence of earthquake precursory phenomena in the 17 January 1995
Kobe earthquake, Japan,'' The European Physical Journal B --Condensed
Matter and Complex Systems \textbf{15}, 551--555 (2000).

\bibitem{Feigenbaum} J. A. Feigenbaum, P. G. O. Freund, ``Discrete
scaling in stock markets before crashes,'' Int. J. Mod. Phys. \textbf{B10},
3737--3745 (1996).

\bibitem{Gluzman} S. Gluzman, V. I. Yukalov, ``Booms and crashes
in self-similar markets,'' Modern Physics Letters\textbf{ B12}, 575--587
(1998).

\bibitem{Drozdz} S. Drozdz, F. Ruf, J. Speth, M. Wojcik, ``Imprints
of log-periodic self-similarity in the stock market,'' European Physical
Journal \textbf{10}, 589--593 (1999).

\bibitem{Sornette} D. Sornette, ``Critical market crashes,'' Physics
reports \textbf{378} (1), 1--98 (2003).

\bibitem{Saleur_1} H. Saleur, C. G. Sammis and D. Sornette, ``Discrete
scale invariance, complex fractal dimensions and log-periodic corrections
in earthquakes,'' J. Geophys. Res. \textbf{101}, 17661--17677 (1996).

\bibitem{Saleur_2} H. Saleur, D. Sornette, ``Complex exponents and
log-periodic corrections in frustrated systems,'' Journal de Physique
I \textbf{6} (3), 327--355 (1996).

\bibitem{MKJ} R. Metzler, J. Klafter and J. Jortner, ``Hierarchies
and logarithmic oscillations in the temporal relaxation patterns of
proteins and other complex systems,'' Proc. Natl. Acad. Sci. USA
\textbf{96}, 11085--11089 (1999).

\bibitem{RTV} R. Rammal, G. Toulose and M. A. Virasoro, ``Ultrametricity
for physicists,'' Rev. Mod. Phys. \textbf{58} (3), 765--788 (1986).

\bibitem{Dayson} F. J. Dayson, ``Existence of a phase-transition
in a one-dimentional Ising ferromagnet,'' Commun. Math. Phys. \textbf{12}
, 91--107 (1969).

\bibitem{Parisi1} G. Parisi, ``Infinite number of order parameters
for spin-glasses,'' Phys. Rev. Lett. \textbf{43} 1754--1756 (1979).

\bibitem{Dot} V. Dotsenko, \emph{An Introduction to the spin glasses
and neural networks} (World Scientific Publishing, Singapore, 1994).

\bibitem{Frauen1} H. Frauenfelder, ``The connection between low-temperature
kinetics and life,'' in: R. H. Austin et al. (Eds.), \emph{Protein
Structure, Molecular and Electronic Reactivity}, 245--261 (Springer,
New York, 1987).

\bibitem{Frauen2} H. Frauenfelder, ``Complexity in proteins,''
Nature Struct. Biol. \textbf{2}, 821--823 (1995).

\bibitem{ALL} B. Dragovich, A. Yu. Khrennikov, S. V. Kozyrev and
I. V. Volovich, ``On $p$-adic mathematical physics,'' $p$-Adic Num.
Ultrametr. Anal. Appl. \textbf{1} (1), 1--17, (2009).

\bibitem{ALL_1} B. Dragovich, A. Yu. Khrennikov, S. V. Kozyrev, I.
V. Volovich and E. I. Zelenov, ``$p$-Adic mathematical physics: The
first 30 years,'' $p$-Adic Num. Ultrametr. Anal. Appl. \textbf{9}
(2), 87--121, (2017).

\bibitem{ALL_2} B. Dragovich, A. Y. Khrennikov, S. V. Kozyrev and
N. \v{Z}. Mi\v{s}i\'{c}, ``$p$-Adic mathematics and theoretical biology,''
Biosystems \textbf{199}, 104288--104288 (2021).

\bibitem{ABK_1999} V. A. Avetisov, A. Kh. Bikulov and S. V. Kozyrev,
``Application of $p$-adic analysis to models of spontaneous breaking
of replica symmetry,'' J. Phys. A:Math. Gen.\textbf{ 32} (50), 8785--8791
(1999).

\bibitem{ABKO_2002} V. A. Avetisov, A. Kh. Bikulov, S. V. Kozyrev
and V. A. Osipov, ``$p$-Adic Models of ultrametric diffusion constrained
by hierarchical energy landscapes,'' J. Phys. A:Math. Gen. \textbf{35}
(2), 177--189 (2002).

\bibitem{ABO_2003} V. A. Avetisov, A. Kh. Bikulov and V. A. Osipov,
``$p$-Adic description of characteristic relaxation in complex systems,''
J. Phys. A:Math. Gen. \textbf{36} (15), 4239--4246 (2003).

\bibitem{ABO_2004} V. A. Avetisov, A. Kh. Bikulov and V. A. Osipov,
``$p$-Adic models for ultrametric diffusion in conformational dynamics
of macromolecules,'' Tr. Mat. Inst. Steklova\textbf{ 245}, 55--64
(2004).

\bibitem{AB_2008} V. A. Avetisov, and A. Kh. Bikulov, ``Protein ultrametricity
and spectral diffusion in deeply frozen proteins,'' Biophys. Rev. Lett.
\textbf{3}, 387--396 (2008).

\bibitem{ABZ_2009} V. A. Avetisov, A. Kh. Bikulov and A. P. Zubarev,
``First passage time distribution and number of returns for ultrametric
random walk,'' J. Phys. A:Math. Gen. \textbf{42} (8), 85005--85021
(2009).

\bibitem{ABZ_2011} V. A. Avetisov, A. Kh. Bikulov and A. P. Zubarev,
``Mathematical modeling of molecular ``nano-machines'','' Vestn. Samar.
Gos. Tekhn. Univ. Ser. Fiz.-Mat. Nauki \textbf{1} (22), 9--15 (2011).

\bibitem{ABZ_2013} V. A. Avetisov, A. Kh. Bikulov and A. P. Zubarev,
``Ultrametricity as a basis for organization of protein molecules:
CO binding to myoglobin,'' Vestn. Samar. Gos. Tekhn. Univ. Ser. Fiz.-Mat.
Nauki \textbf{1} (30), 315-- 325 (2013).

\bibitem{ABZ_2014} V. A. Avetisov, A. Kh. Bikulov and A. P. Zubarev,
``Ultrametric random walk and dynamics of protein molecule,'' Tr.
Mat. Inst. Steklova \textbf{285}, 9--32 (2014).

\bibitem{BZ_2021} A. Kh. Bikulov and A. P. Zubarev, ``Ultrametric
theory of conformational dynamics of protein molecules in a functional
state and the description of experiments on the kinetics of CO binding
to myoglobin,'' Physica A: Statistical Mechanics and its Applications
\textbf{583}, 126280--126280 (2021).

\bibitem{Stillinger1} F. H. Stillinger, T. A. Weber, ``Dynamics
of structural transitions in liquids,'' Phys. Rev. \textbf{A 25},
2408--2416 (1982).

\bibitem{Stillinger2} F. H. Stillinger, T. A. Weber, ``Packing structures
and transitions in liquids and solids,'' Science \textbf{225}, 983--989
(1984).

\bibitem{BK} O. M. Becker, M. Karplus, ``The topology of multidimensional
protein energy surfaces: Theory and application peptide structure
and kinetics,'' J. Chem. Phys. \textbf{106}, 1495--1517 (1997).

\bibitem{Koz} S.V. Kozyrev, ``Methods and applications of ultrametric
and $p$-adic analysis: From wavelet theory to biophysics,'' Proc.
Steklov Inst. Math. \textbf{274} (1), 1--84 (2011).

\bibitem{VVZ} V. S. Vladimirov, I. V. Volovich and E. I. Zelenov,
\emph{$p$-Adic analysis and mathematical physics} (World Scientific
Publishing, Singapore, 1994).

\bibitem{BZ_2022} A. Kh. Bikulov and A. P. Zubarev, ``The sojourn
time problem for a $p$-Adic random walk and its applications,''
$p$-Adic Numbers, Ultrametric Analysis and Applications \textbf{14}
(4) , 265--278 (2022).

\bibitem{PFVBB} V. V. Ponkratov, J. Friedrich, K. M. Vanderkooi,
A. L. Burin and Yu. A. Berlin, ``Physics of protein at low temperature,''
J. Low. Temp. Phys. \textbf{3}, 289--317 (2006).

\bibitem{Friedrich1} J. Schlichter, J. Friedrich, L. Herenyi and
J. Fidy, ``Protein dynamics at low temperatures,'' The Journal of
Chemical Physics \textbf{112} (6), 3045--3050 (2000).

\bibitem{Friedrich2} J. Schlichter, K. D. Fritsch, J. Friedrich,
and J. M. Vanderkooi, ``Conformational dynamics of a low temperature
protein: Free base cytochrome-c,'' The Journal of Chemical Physics
\textbf{110}, 3229--3234 (1999).

\bibitem{ABB} A. Ansary, J. Berendzen, S.E. Boune, et al., ``Protein
states and proteinquake,'' Proc. Natl. Acad. Sci. USA \textbf{82},
5000--5004 (1985).

\bibitem{SAB} P. J. Stenbach, A. Ansary, J. Berendzen, et al, ``Ligand
binding to heme proteins: Connection between dinamics and function,''
Biochemistry \textbf{30}, 3988--4001 (1991).

\bibitem{Bik_2010} A. K. Bikulov, ``On solution properties of some
types of $p$-adic kinetic equations of the form reaction-diffusion,''
$p$-Adic Numbers, Ultrametric Analysis and Applications \textbf{2}
(3), 187--206 (2010).

\bibitem{LJA_1993} M. Lim, T. A. Jackson and P. A. Anfinrud, ``Nonexponential
protein relaxation: dynamics of conformational change in myoglobin,''
Proceedings of the National Academy of Sciences \textbf{90} (12),
5801--5804 (1993).

\bibitem{JLA_1994} T. A. Jackson, M. Lim and P. A. Anfinrud, ``Complex
nonexponential relaxation in myoglobin after photodissociation of
MbCO: measurement and analysis from 2 ps to 56 us,'' Chemical physics
\textbf{189} (2-3), 131--140 (1994).

\bibitem{Frau_M} H. Frauenfelder and B. H. McMahon, ``Energy landscape
and fluctuations in proteins, '' Annalen der Physik\textbf{ 9} (9-10),
655--667 (2000).

\bibitem{Sabelko_1} J. Sabelko, J. Ervin and M. Gruebele, ``Observation
of strange kinetics in protein folding,'' Proceedings of the National
Academy of Sciences, \textbf{96} (11), 6031--6036 (1999)

\bibitem{Sabelko_2} S. Osv\'{a}th, J. J. Sabelko and M. Gruebele,
``Tuning the heterogeneous early folding dynamics of phosphoglycerate
kinase,'' Journal of molecular biology \textbf{333} (1), 187--199
(2003).

\bibitem{Remmert} R. Remmert, \emph{Classical topics in complex function
theory} (Springer Science \& Business Media, Vol. 172, 2013).
\end{thebibliography}
\end{document}